\documentclass[runningheads]{llncs}
\usepackage[hidelinks]{hyperref}
\usepackage[font=scriptsize]{caption}
\usepackage{graphicx}
\usepackage{diagbox}
\usepackage{comment}
\usepackage{listings}

\begin{document}
\title{PM4Py-GPU: a High-Performance General-Purpose Library for Process Mining}
\titlerunning{PM4Py-GPU}
%
\author{
Alessandro Berti\inst{1,2}
\and
Minh Phan Nghia
\and
Wil M.P. van der Aalst\inst{1,2}
}
\authorrunning{A. Berti et al.}
%
\institute{Process and Data Science Group @ RWTH Aachen, Aachen, Germany
\email{\{a.berti, wvdaalst\}@pads.rwth-aachen.de, minh.nghia.phan@rwth-aachen.de }\\
\and
Fraunhofer Institute of Technology (FIT), Sankt Augustin, Germany\\}
%
\maketitle              
\begin{abstract}
Open-source process mining provides many algorithms for the analysis of event data which could be used to analyze mainstream processes (e.g., O2C, P2P, CRM).
However, compared to commercial tools, they lack the performance and struggle to analyze large amounts of data.
This paper presents PM4Py-GPU, a Python process mining library based on the NVIDIA RAPIDS framework.
Thanks to the dataframe columnar storage and the high level of parallelism, a significant speed-up is achieved on classic process mining computations and processing activities.
\end{abstract}


\section{Introduction}
\label{sec:introduction}

Process mining is a branch of data science that aims to analyze the execution of business processes starting from the event data contained in the information
systems supporting the processes. Several types of process mining are available, including \emph{process discovery} (the automatic discovery of a process model
from the event data), \emph{conformance checking} (the comparison between the behavior contained in the event data against the process model, with the purpose to
find deviations), \emph{model enhancement} (the annotation of the process model with frequency/performance information) and \emph{predictive analytics}
(predicting the next path or the time until the completion of the instance).
Process mining is applied worldwide to a huge amount of data using different tools (academic/commercial). Some important tool features to allow process mining in organizational settings
are: the \emph{pre-processing and transformation} possibilities,
the \emph{possibility to drill-down} (creating smaller views on the dataset, to focus on some aspect of the process), the availability of \emph{visual analytics}
(which are understandable to non-business users), the \emph{responsiveness and performance} of the tool, and the possibilities of \emph{machine learning}
(producing useful predictive analytics and what-if analyses). Commercial tools tackle these challenges with more focus than academic/open-source tools,
which, on the other hand, provide more complex analyses (e.g., process discovery with inductive miner, declarative conformance checking).
The PM4Py library \url{http://www.pm4py.org}, based on the Python 3 programming language, permits to integrate with the data processing and machine learning packages which are
available in the Python world (Pandas, Scikit-Learn). However, most of its algorithms work in single-thread, which is a drawback for performance.
In this demo paper, we will present a GPU-based open-source library for process mining, PM4Py-GPU, based on the NVIDIA RAPIDS framework, 
allowing us to analyze a large amount of event data with high performance offering access to GPU-based machine learning. The speedup over other open-source libraries
for general process mining purposes is more than 10x.
The rest of the demonstration paper is organized as follows. Section \ref{sec:preliminaries} introduces the NVIDIA RAPIDS framework, which is at the base of PM4Py-GPU, and
of some data formats/structures for the storage of event logs;
Section \ref{sec:implementationTool} presents the implementation, the different components of the library and some code examples; Section \ref{sec:assessment} assess PM4Py-GPU against other products;
Section \ref{sec:relatedWork} introduces the related work on process mining on big data and process mining on GPU; Finally, Section \ref{sec:conclusion} concludes the demo paper.

\section{Preliminaries}
\label{sec:preliminaries}

This section will first present the NVIDIA RAPIDS framework for GPU-enabled data processing and mining. Then, an overview of the most widely used file formats and data structures
for the storage of event logs is provided.

\subsection{NVIDIA RAPIDS}

The NVIDIA RAPIDS framework \url{https://developer.nvidia.com/rapids} was launched by NVIDIA in 2018 with the purpose to enable general-purpose data science pipelines
directly on the GPU. It is composed of different components: CuDF (GPU-based dataframe library for Python, analogous to Pandas), CuML (GPU-based general-purpose machine learning library for Python,
similar to Scikit-learn), and CuGraph (GPU-based graph processing library for Python, similar to NetworkX). The framework is based on CUDA (developed by NVIDIA to allow low-level programming on the GPU)
and uses RMM for memory management.
NVIDIA RAPIDS exploit all the cores of the GPU in order to maximize the throughput.
When a computation such as retrieving the maximum numeric value of a column is operated against a column of the dataframe, the different cores of the GPU act on different parts of the column,
a maximum is found on every core. Then the global maximum is a reduction of these maximums. Therefore, the operation is parallelized on all the cores of the GPU.
When a group-by operation is performed, the different groups are identified (also here using all the cores of the GPU) as the set of rows indices. Any operation on the group-by operation
(such as taking the last value of a column per group; performing the sum of the values of a column per group; or calculating the difference between consecutive values in a group)
is also performed exploiting the parallelism on the cores of the GPU.

\subsection{Dataframes and File Formats for the Storage of Event Logs}

In this subsection, we want to analyze the different file formats and data structures that could be used to store event logs,
and the advantages/disadvantages of a columnar implementation.
As a standard to interchange event logs, the XES standard is proposed \url{https://xes-standard.org/}, which is text/XML based.
Therefore, the event log can be ingested in memory after parsing the XML, and this operation is quite expensive.
Every attribute in a XES log is typed, and the attributes for a given case do not need to be replicated among all the events.
Event logs can also be stored as CSV(s) or Parquet(s), both resembling the structure of a table.
A CSV is a textual file hosting an \emph{header row} (containing the names of the different columns separated by a separator character)
and many \emph{data rows} (containing the values of the attributes for the given row separated by a separator character).
A problem with the CSV format is the typing of the attributes.
A Parquet file is a binary file containing the values for each column/attribute, and applying a column-based compression. Each column/attribute
is therefore strictly typed.
CuDF permits the ingestion of CSV(s)/Parquet(s) into a dataframe structure.
A dataframe is a table-like data structure organized as columnar storage. As many data processing operations work on a few attributes/columns
of the data, adopting a columnar storage permits to retrieve specific columns with higher performance and to reduce performance problems such as cache misses.
Generally, the ingestion of a Parquet file in a CuDF dataframe is faster because the data is already organized in columns. In contrast, the parsing of the text of a CSV and its transformation to a dataframe is
more time expensive. However, NVIDIA CuDF is also impressive in the ingestion of CSV(s) because the different cores of the GPU are used on different parts of the CSV file.

\section{Implementation and Tool}
\label{sec:implementationTool}

In PM4Py-GPU, we assume an event log to be ingested from a Parquet/CSV file into a CuDF dataframe using the methods available in CuDF.
On top of such dataframe, different operations are possible, including:
\begin{itemize}
\item \emph{Aggregations/Filtering at the Event Level}: we would like to filter in/out a row/event or perform any aggregation based solely on the properties of the row/event. Examples:
filtering the events/rows for which the cost is $> 1000$; associate its number of occurrences to each activity.
\item \emph{Aggregations/Filtering at the Directly-Follows Level}: we would like to filter in/out rows/events or perform any sort of aggregation based on the properties of the event and of the previous (or next) event. Examples:
filtering the events with activity \emph{Insert Fine Notification} having a previous event with activity \emph{Send Fine}; calculating the frequency/performance directly-follows graph.
\item \emph{Aggregations/Filtering at the Case Level}: this can be based on global properties of the case (e.g., the number of events in the case or the throughput time of the case) or on properties of the single event. In this setting, we need an initial exploration of the dataframe to group the indexes of the rows based on their case and then perform the filtering/aggregation on top
of it. Examples: filtering out the cases with more than $10$ events; filtering the cases with at least one event with activity \emph{Insert Fine Notification}; finding the throughput time for all the cases of the log.
\item \emph{Aggregations/Filtering at the Variant Level}: the aggregation associates each case to its variant. The filtering operation accepts a collection of variants and keeps/remove all the cases whose variant
fall inside the collection. This requires a double aggregation: first, the events need to be grouped in cases. Then this grouping is used to aggregate the cases into the variants.
\end{itemize}

To facilitate these operations, in PM4Py-GPU we operate three steps starting from the original CuDF dataframe:
\begin{itemize}
\item The dataframe is ordered based on three criteria (in order, case identifier, the timestamp, and the absolute index of the event in the dataframe), to have the events of the same cases near each other 
in the dataframe, increasing the efficiency of group-by operations.
\item Additional columns are added to the dataframe (including the position of the event inside a case; the timestamp and the activity of the previous event) to allow for aggregations/filtering at the
directly-follows graph level.
\item A \emph{cases dataframe} is found starting from the original dataframe and having a row for each different case in the log. The columns of this dataframe include the number of events for the case,
the throughput time of the case, and some numerical features that uniquely identify the case's variant. Case-based filtering is based on both the original dataframe and the cases dataframe. Variant-based filtering
is applied to the cases dataframe and then reported on the original dataframe (keeping the events of the filtered cases).
\end{itemize}

The PM4Py-GPU library is available at the address \url{https://github.com/Javert899/pm4pygpu}. It does not require any further dependency than the NVIDIA RAPIDS library, which by itself depends on
the availability of a GPU, the installation of the correct set of drivers, and of NVIDIA CUDA. The different modules of the library are:
\begin{itemize}
\item \emph{Formatting module (format.py)}: performs the operations mentioned above on the dataframe ingested by CuDF. This enables the rest of the operations described below.
\item \emph{DFG retrieval / Paths filtering (dfg.py)}: discovers the frequency/performance directly-follows graph on the dataframe. This enables paths filtering on the dataframe.
\item \emph{EFG retrieval / Temporal Profile (efg.py)}: discovers the eventually-follows graphs or the temporal profile from the dataframe.
\item \emph{Sampling (sampling.py)}: samples the dataframe based on the specified amount of cases/events.
\item \emph{Cases dataframe (cases\_df.py)}: retrieves the cases dataframe. This permits the filtering on the number of events and on the throughput time.
\item \emph{Variants (variants.py)}: enables the retrieval of variants from the dataframe. This permits variant filtering.
\item \emph{Timestamp (timestamp.py)}: retrieves the timestamp values from a column of the dataframe. This permits three different types of timestamp filtering (events, cases contained, cases intersecting).
\item \emph{Endpoints (start\_end\_activities.py)}: retrieves the start/end activities from the dataframe. This permits filtering on the start and end activities.
\item \emph{Attributes (attributes.py)}: retrieves the values of a string/numeric attribute. This permits filtering on the values of a string/numeric attribute.
\item \emph{Feature selection (feature\_selection.py)}: basilar feature extraction, keeping for every provided numerical attribute the last value per case, and for each provided string attribute its one-hot-encoding.
\end{itemize}
An example of usage of the PM4Py-GPU library, in which a Parquet log is ingested, and the directly-follows graph is computed, is reported in the following listing.

\begin{table*}[!t]
\caption{Event logs used in the assessment, along with their number of events, cases, variants and activities.}
\centering
\resizebox{0.83\textwidth}{!}{%
\begin{tabular}{|l|l|l|l|l|}
\hline
{\bf Log} & {\bf Events} & {\bf Cases} & {\bf Variants} & {\bf Activities} \\
\hline
roadtraffic\_2 & 1,122,940 & 300,740 & 231 & 11 \\
roadtraffic\_5 & 2,807,350 & 751,850 & 231 & 11 \\
roadtraffic\_10 & 5,614,700 & 1,503,700 & 231 & 11 \\
roadtraffic\_20 & 11,229,400 & 3,007,400 & 231 & 11 \\
\hline
bpic2019\_2 & 3,191,846 & 503,468 & 11,973 & 42 \\
bpic2019\_5 & 7,979,617 & 1,258,670 & 11,973 & 42 \\
bpic2019\_10 & 15,959,230 & 2,517,340 & 11,973 & 42 \\
\hline
bpic2018\_2 & 5,028,532 & 87,618 & 28,457 & 41 \\
bpic2018\_5 & 12,571,330 & 219,045 & 28,457 & 41 \\
bpic2018\_10 & 25,142,660 & 438,090 & 28,457 & 51 \\
\hline
\end{tabular}
}
\label{tab:eventLogData}
\end{table*}

\begin{lstlisting}[label=lst:example]
import cudf
from pm4pygpu import format, dfg
df = cudf.read_parquet('receipt.parquet')
df = format.apply(df)
frequency_dfg = dfg.get_frequency_dfg(df)
\end{lstlisting}
\captionof{lstlisting}{Example code of PM4Py-GPU.}

\section{Assessment}
\label{sec:assessment}

In this section, we want to compare PM4Py-GPU against other libraries/solutions for process mining to evaluate mainstream operations' execution time against
significant amounts of data. The compared solutions include PM4Py-GPU (described in this paper), PM4Py (CPU single-thread library for process mining in Python; \url{https://pm4py.fit.fraunhofer.de/}), the PM4Py Distributed Engine (described in the assessment).
All the solutions have been run on the same machine (Threadripper 1920X, 128 GB of DDR4 RAM, NVIDIA RTX 2080).
The event logs of the assessment include the Road Traffic Fine Management \url{https://data.4tu.nl/articles/dataset/Road_Traffic_Fine_Management_Process/12683249},
the BPI Challenge 2019 \url{https://data.4tu.nl/articles/dataset/BPI_Challenge_2019/12715853} and the BPI Challenge 2018 \url{https://data.4tu.nl/articles/dataset/BPI_Challenge_2018/12688355}
event logs. The cases of every one of these logs have been replicated $2$, $5$, and $10$ times for the assessment (the variants and activities are unchanged). Moreover, the smallest of these
logs (Road Traffic Fine Management log) has also been replicated $20$ times.
The information about the considered event logs is reported in Table \ref{tab:eventLogData}. In particular, the suffix (\_2, \_5, \_10) indicates the number of replications of the cases of the log.
The results of the different experiments is reported in Table \ref{tab:expExecutionTime}. The first experiment is on the importing time (PM4Py vs. PM4Py-GPU; the other two software cannot be
directly compared because of more aggressive pre-processing). We can see that PM4Py-GPU is slower than PM4Py in this setting (data in the GPU is stored in a way that facilitates parallelism).
The second experiment is on the computation of the directly-follows graph in the four different platforms. Here, PM4Py-GPU is incredibly responsive
The third experiment is on the computation of the variants in the different platforms. Here, PM4Py-GPU and the PM4Py Distributed Engine perform both well
(PM4Py-GPU is faster to retrieve the variants in logs with a smaller amount of variants).

\begin{table}[ht]
\caption{Comparison between the execution times of different tasks. The configurations analyzed are: P4 (single-core PM4Py), P4G (PM4Py-GPU), P4D (PM4Py Distributed Engine).
The tasks analyzed are: importing the event log from a Parquet file, the computation of the DFG and the computation of the variants. For the PM4Py-GPU (computing the DFG and variants), the speedup
in comparison to PM4Py is also reported.}
\resizebox{\textwidth}{!}{%
\begin{tabular}{|l|cc|ccc|ccc|}
\hline
~ & \multicolumn{2}{|>{}c|}{Importing} & \multicolumn{3}{|>{}c|}{DFG} & \multicolumn{3}{|>{}c|}{Variants} \\
{\bf Log} & {\bf P4} & {\bf P4G} & {\bf P4} & {\bf P4G} & {\bf P4D} & {\bf P4} & {\bf P4G} & {\bf P4D} \\
\hline
roadtraffic\_2 & {\bf 0.166s} & 1.488s & 0.335s & {\bf 0.094s} (3.6x) & 0.252s & 1.506s & {\bf 0.029s} (51.9x) & 0.385s \\
roadtraffic\_5 & {\bf 0.375s} & 1.691s & 0.842s & {\bf 0.098s} (8.6x) & 0.329s & 3.463s & {\bf 0.040s} (86.6x) & 0.903s \\
roadtraffic\_10 & {\bf 0.788s} & 1.962s & 1.564s & {\bf 0.105s} (14.9x) & 0.583s & 7.908s & {\bf 0.055s} (144x) & 1.819s \\
roadtraffic\_20 & {\bf 1.478s} & 2.495s & 3.200s & {\bf 0.113s} (28.3x) & 1.048s & 17.896s & {\bf 0.092s} (195x) & 3.380s \\
\hline
bpic2019\_2 & {\bf 0.375s} & 1.759s & 0.980s & {\bf 0.115s} (8.5x) & 0.330s & 3.444s & 0.958s (3.6x) & {\bf 0.794s} \\
bpic2019\_5 & {\bf 0.976s} & 2.312s & 2.423s & {\bf 0.156s} (15.5x) & 0.613s & 8.821s & {\bf 0.998s} (8.9x) & 1.407s \\
bpic2019\_10 & {\bf 1.761s} & 3.156s & 4.570s & {\bf 0.213s} (21.5x) & 1.679s & 19.958s & {\bf 1.071s} (18.6x) & 4.314s \\
\hline
bpic2018\_2 & {\bf 0.353s} & 1.846s & 1.562s & {\bf 0.162s} (9.6x) & 0.420s & 6.066s & 5.136s (1.2x) & {\bf 0.488s} \\
bpic2018\_5 & {\bf 0.848s} & 2.463s & 3.681s & {\bf 0.214s} (17.2x) & 0.874s & 14.286s & 5.167s (2.8x) & {\bf 0.973s} \\
bpic2018\_10 & {\bf 1.737s} & 3.470s & 7.536s & {\bf 0.306s} (24.6x) & 1.363s & 29.728s & 5.199s (5.7x) & {\bf 1.457s} \\
\hline
\end{tabular}
}
\label{tab:expExecutionTime}
\end{table}

\section{Related Work}
\label{sec:relatedWork}

{\bf Process Mining on Big Data Architectures}:
an integration between process mining techniques and Apache Hadoop has been proposed in \cite{DBLP:conf/bpm/HernandezZEA15}. Apache Hadoop does not work in-memory and requires the serialization of every step.
Therefore, technologies such as Apache Spark could be used for in-memory process mining\footnote{\url{https://www.pads.rwth-aachen.de/go/id/ezupn/lidx/1}}. The drawback of Spark is the additional overhead
due to the log distribution step, which limits the performance benefits of the platform. Other platform such as Apache Kafka have been used for processing of streams \cite{DBLP:journals/sncs/NogueiraR21}.
Application-tailored engines have also been proposed. The ``PM4Py Distributed engine'' \footnote{\url{https://www.pads.rwth-aachen.de/go/id/khbht}} has been proposed as a multi-core and multi-node engine
tailored for general-purpose process mining with resource awareness. However, in contrast to other distributed engines, it misses any failure-recovery option and therefore is not good for very long lasting computations. The Process Query Language (PQL) is integrated in the Celonis commercial process mining software \url{https://www.celonis.com/}
and provides high throughput for mainstream process mining computations in the cloud. \\
{\bf Data/Process Mining on GPU}:
many popular data science algorithms have been implemented on top of a GPU \cite{DBLP:journals/widm/Cano18}. In particular, the training of machine learning models, which involve tensor operations, can
have huge speed-ups using the GPU rather than the CPU. In \cite{DBLP:conf/caise/TaxVRD17} (LSTM neural networks) and \cite{DBLP:conf/icpm/Pasquadibisceglie19} (convolutional neural networks),
deep learning approaches are used for predictive purposes.
Some of the process mining algorithms have been implemented on top of a GPU. In \cite{DBLP:conf/bpm/KundraJS15}, the popular alpha miner algorithm is implemented on top of GPU and compared against
the CPU counterpart, showing significant gains. In \cite{DBLP:conf/bpm/FerreiraS16}, the discovery of the paths in the log is performed on top of a GPU with a big speedup in the experimental setting.

\section{Conclusion}
\label{sec:conclusion}

In this paper, we presented PM4Py-GPU, a high-performance library for process mining in Python, which is based on the NVIDIA RAPIDS framework for GPU computations. The experimental results against
distributed open-source software (PM4Py Distributed Engine) are very good, and the library seems suited for process mining on a significant amount of data.
However, an expensive GPU is needed to make the library work, which could be a drawback for widespread usage.
We should also say that the number of process mining functionalities supported by the GPU-based library is limited, hence comparisons against open-source/commercial software supporting a more comprehensive number of features might be unfair.

\section*{Acknowledgements}

We thank the Alexander von Humboldt (AvH) Stiftung for supporting our research.

\bibliographystyle{splncs04}
\bibliography{pm4pygpu}

\end{document}